# Shadows of naked singularity in Brans–Dicke gravity

**Prajwal Hassan Puttasiddappa**[1,2,a], **Davi C. Rodrigues**[3,4], **David F. Mota**[2]

[1] PPGCosmo, Núcleo de Astrofísica e Cosmologia (Cosmo-Ufes), Universidade Federal do Espírito Santo, Vitória, ES 29075-910, Brazil
[2] Institute of Theoretical Astrophysics, University of Oslo, Sem Sælands vei 13, 0371 Oslo, Norway
[3] Departamento de Física, Núcleo de Astrofísica e Cosmologia (Cosmo-Ufes), PPGCosmo, Universidade Federal do Espírito Santo, Vitória, ES 29075-910, Brazil
[4] Centro Brasileiro de Pesquisas Físicas (CBPF), R. Xavier Sigaud 150, Rio de Janeiro, RJ 22290-180, Brazil

**Abstract** We investigate exact vacuum solutions in Brans–Dicke (BD) gravity, focusing on their implications for black hole shadow imaging. Motivated by the Event Horizon Telescope (EHT) observations, we revisit a class of BD solutions that exhibit a naked singularity. These solutions, despite lacking a conventional event horizon, exhibit photon spheres and produce shadow-like features. We analyze null geodesics and perform ray-tracing simulations under an optically thin accretion disk model to generate synthetic images. Our results show that BD naked singularities can cast shadows smaller than those of Schwarzschild black holes of equivalent mass. This requires, however, negative values for the BD parameter $\omega$ (in the range $-3/2 < \omega < 0$). The lower bound on $\omega$ is compatible with the absence of ghosts. Although such $\omega$ range is outside the Solar System bounds, our results are relevant for a larger class of modified gravity models whose local behaviour can be approximated by Brans–Dicke gravity. These findings suggest that BD naked singularities are possible candidates for compact astrophysical objects.

## 1 Introduction

Significant progress has recently been made in testing gravity in the strong-field regime on three main fronts: (i) Precise measurements of stellar orbits near the Galactic Center [1–6]; (ii) direct detection of gravitational waves [7–9]; and (iii) black hole (BH) observations of the Event Horizon Telescope (EHT). The EHT observations include highly lensed emissions near supermassive BHs M87* [10] and Sgr A* [11], often referred to as "BH images". They feature the "black hole shadow" which broadly refers to the brightness depression at the center. These images are commonly interpreted as highly lensed images of the BH accretion disk. The brightness depression is expected to result from extreme light bending and light capture by the central supermassive BH. In a stricter sense, the "black hole shadow" refers to the region from which no photons can escape, being delimited by the photon ring. For a detailed discussion of the various terminologies used in this context, see [12,13].

Current observations do not directly constrain the photon ring radius, as it remains too faint for the EHT to resolve. The observed photon intensity is dominated by the accretion disk, which serves as the primary photon source illuminating the central BH. However, the size of the BH shadow in EHT images, particularly for Sgr A*, can be used to place constraints on modified gravity [14].

There is currently a considerable effort focused on testing gravity models beyond General Relativity (GR), driven by a diverse range of motivations. These include addressing anomalies in cosmological data, providing alternatives to the yet-undiscovered dark matter and dark energy, and incorporating quantum corrections, among others. Furthermore, testing gravity beyond GR provides useful insight for understanding GR itself.

We focus on Brans–Dicke (BD) gravity [15,16] due to its simplicity, as it introduces only a single scalar field ($\psi$) and a single parameter ($\omega$), and several modified gravity theories reduce to BD in appropriate limits. The general picture we consider is that the true gravity theory could be different from both GR and BD, but locally it could resemble either one of these cases. For instance, while the Solar System is well-approximated by GR, or BD with $\omega \gtrsim 40\,000$ [17–19], distant regions could contain *black-hole–like* compact objects (e.g., stationary naked-singularity solutions), whose exterior spacetime is well approximated by BD with different (and possibly negative) effective values of $\omega$. Such scenarios arise naturally in scalar–tensor and higher-derivative theories: the scalar field is an independent degree of freedom that can violate the strong equivalence principle, so its background value and gradients need not be universal, allowing different local structures to experience different effective couplings (e.g., [20–22]).

[a] e-mail: prajwal.puttasiddappa@edu.ufes.br (corresponding author)

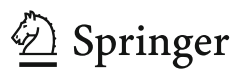

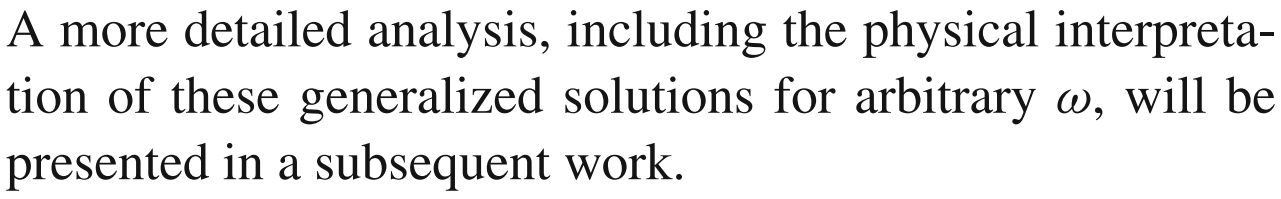

Despite its simplicity, Brans–Dicke (BD) gravity yields several unexpected results. The exact vacuum solutions originally proposed by Brans [16] have been extensively discussed and debated in the literature [23–27], and have been revisited in greater detail in [28]. In addition to these solutions, other noteworthy proposals, such as those in [29,30] have received comparatively less attention. Our work revisits them.

We show in Appendix A that the solution presented in [29] extends the solutions with solitonic configuration previously studied in [30] to arbitrary values of $\omega$. The solution in [30] was originally derived for the specific case $\omega = 0$. In this work, we focus exclusively on the exact solution from [29], with a particular focus on the observational signatures, such as the black hole shadow.

## 2 Shadows of Brans–Dicke vacuum solution

BD action is given by:

$$S = \frac{1}{16\pi} \int d^4x \sqrt{-g} \left( \psi R - \frac{\omega}{\psi} (\nabla \psi)^2 \right) , \tag{1}$$

and the vacuum field equations are,

$$G_{\mu\nu} = \frac{\omega}{\psi^2} \left( \nabla_\mu \psi \nabla_\nu \psi - \frac{1}{2} g_{\mu\nu} (\nabla \psi)^2 \right) + \frac{\nabla_\mu \nabla_\nu \psi}{\psi} , \qquad \Box \psi = 0 . \tag{2}$$

With the static, spherically symmetric metric ansatz,

$$ds^2 = -e^{A(r)}\, dt^2 + e^{B(r)}\, dr^2 + r^2 (d\theta^2 + \sin^2\theta\, d\varphi^2) , \tag{3}$$

we solve the field equations to find, for $\omega \neq -2$ [29]:

$$e^{A(r)} = \left( \frac{\sqrt{1 + \frac{r_*^2}{r^2}} - \frac{r_*}{r}}{\sqrt{1 + \frac{r_*^2}{r^2}} + \frac{r_*}{r}} \right)^{\frac{1}{\beta}} , \quad e^{B(r)} = \frac{1}{1 + \frac{r_*^2}{r^2}}, \tag{4}$$

where $r_* = \beta G M$, with $\beta = \sqrt{(\omega+2)/2}$. In the following, we restrict our attention to cases where $r_*^2 > 0$, or equivalently, $\beta^2 > 0$, which corresponds to gravity remaining an attractive interaction. There is a true singularity at $r = 0$, that can be verified by computing the Kretschmann scalar. However, there is no event horizon protecting this singularity.

In Appendix A, we briefly demonstrate that the solution (4) represents a generalization of the class of solutions obtained in [30], where the BD parameter was fixed to $\omega = 0$. A more detailed analysis, including the physical interpretation of these generalized solutions for arbitrary $\omega$, will be presented in a subsequent work.

### 2.1 Photon geodesics in general spherically symmetric spacetime

Given the general spherically symmetric metric (3), we investigate the geodesics by restricting to the equatorial plane, $\theta = \pi/2$. The symmetries of this metric yield two conserved quantities: the energy per unit mass $E = e^{A(r)} \dot{t}$ and the angular momentum per unit mass $L = r^2 \dot{\varphi}$. The ratio $b = L/E$ defines the impact parameter. The geodesic equation is $-e^{A(r)}\, \dot{t}^2 + e^{B(r)}\, \dot{r}^2 + r^2 \dot{\varphi}^2 = -\varepsilon$ where the overdot denotes differentiation with respect to the affine parameter and $\varepsilon$ is a normalization parameter, with $\varepsilon = 0$ for null vectors and $\varepsilon = 1$ for timelike vectors. We can express the radial geodesic equation as:

$$\left( \frac{e^{A+B}}{L^2} \right) \dot{r}^2 = \frac{1}{b^2} - V(r) - \varepsilon \frac{e^A}{L^2} , \tag{5}$$

where the effective potential is given by $V(r) = \frac{e^{A(r)}}{r^2}$. It is typical to absorb the factor of $L^2$ in the denominator on the left-hand side by redefining the affine parameter. The critical points of $V(r)$ are of special interest. The minimum $dV(r)/dr = 0$, defines the radius of the *critical photon ring* or the *photon sphere* $r_\gamma$, where photons are trapped in unstable circular orbits. For numerical convenience, we can express the radial null geodesic equation (with $\varepsilon = 0$) as a function of the angle $\varphi$:

$$\left( \frac{dr}{d\varphi} \right)^2 = \frac{r^2 e^{-B(r)}}{b^2} \left( r^2 e^{-A(r)} - b^2 \right) . \tag{6}$$

The trajectories, described by the quadratic shape equation above, encounter a turning point, say at $r = R$, where $\frac{dr}{d\varphi}\big|_R = 0$. This implies $b^2(R) = R^2 e^{-A(R)}$. The smallest possible turning point occurs when $R = r_\gamma$, at which the corresponding, $b(r_\gamma)$ is called the *critical impact parameter*. This marks the boundary of the BH's shadow, $r_{\rm sh} \equiv b(r_\gamma)$. Then we can divide photon trajectories into those that get deflected to infinity if $R > r_\gamma$ or fall into the BH if $R < r_\gamma$. In reality, the critical photon ring consists of an infinite sequence of progressively closer sub-rings. As $R$ approaches $r_\gamma$, the number of light orbits (or photon rings) increases indefinitely, with each subsequent ring lying exponentially closer to the previous one. For example, for Schwarzschild spacetime $e^{A(r)} = e^{-B(r)} = 1 - \frac{2M}{r}$, $r_\gamma = 3M$ and $r_{\rm sh} = 3\sqrt{3}M$.

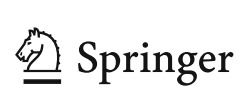

### 2.2 Accretion models and the image

The BH images captured by the EHT reveal a central brightness depression, bordered by a bright feature which supposedly includes the critical photon ring (however, independent data analysis challenges this interpretation [31,32]). The sensitivity limitations of the EHT prevent it from accurately resolving features dimmer than 10% of the peak brightness, leaving the critical photon ring indistinct. Nonetheless, simulated images can reveal finer details, maybe also the substructure of the critical photon ring.

In this work, we perform a basic ray-tracing procedure [33,34] to explore the optical appearance of the BD naked singularity solution. The critical photon ring itself is a geometric feature determined solely by the BH's spacetime geometry. However, its observed appearance also depends on the accretion model illuminating the region [35]. We now consider an accretion disk around the compact object as the light source. Following [34–37], we assume a simplified accretion model that is optically and geometrically thin, avoiding the complexities of magnetohydrodynamics. Optically thin, meaning it does not absorb its own radiation and leaves the specific intensity conserved. The disk is also assumed to source monochromatic radiation corresponding to the frequency measured in the observer's frame. Geometrically, we assume it to be existing as a thin layer confined to the black hole's equatorial plane. For a spherically symmetric disk, a single function $I(r)$ characterizes the intensity profile. We can place the peak of the profile at the innermost stable circular orbit (ISCO) radius, or let it extend down to the event horizon for example. So we define a parameter $\mu$ indicating the peak radius of the intensity profile. The spread of this peak is controlled by a parameter $\sigma$, and the parameter $\gamma$ describes the fall-off of intensity from the peak to infinity. We then have modified Johnson distribution [37]:

$$I(r;\gamma,\mu,\sigma)=\frac{\exp\left\{-\frac{1}{2}\left[\gamma+\operatorname{arcsinh}\left(\frac{r-\mu}{\sigma}\right)^2\right]\right\}}{\sqrt{(r-\mu)^2+\sigma^2}} \tag{7}$$

We adopt this phenomenological form primarily for its analytical tractability and flexibility. The free parameters allow control over the peak location and fall-off behaviour of the emission, enabling us to isolate the geometric influence on shadow features without invoking the full complexity of accretion physics. We set the peak $\mu$ of the distribution at the ISCO radius of massive particles. While models such as the Novikov-Thorne profile [38] or general relativistic magnetohydrodynamic simulations [39,40] can incorporate physical effects like accretion rates, magnetic fields, and radiative processes, our goal here is to examine how the emission geometry alone shapes the observed image, rather than to model the full astrophysical environment.

For the ray tracing procedure itself, we categorize photons reaching an axial observer: we trace each photon back to its turning point near the black hole, grouping them by the number of half-turns ($m$) they make. First, there is the direct emission from the accretion disk $m = 0$, directly to the observer without being lensed by the black hole. These photons contribute significantly to the brightness. The next significant contribution involves those photons that complete a half-turn $m = 1$ around the black hole. These are the lensed photons. For the formation of the photon ring, photons must complete at least $m = 2$ half-turns in which case we have a ring that constitutes the boundary of the observable shadow. In principle, we can continue resolving an infinite sequence of sub-rings for $m > 2$. However, current observations cannot resolve these finer structures due to dominant disk emission. Importantly, $m = 2$ ring is sufficient for testing various gravity models.

## 3 Photon geodesics and ray-tracing around BD naked singularity

From (4), it is clear that the solutions do not exhibit a conventional event horizon. It has a naked singularity at $r = 0$ (assuming $r_*^2 > 0$). The spacetime is asymptotically flat. The time-time component $e^{A(r)}$ remains finite and nonzero for all $r > 0$, vanishing only in the limit $r \to 0$. Consequently, the gravitational redshift of photons emitted at radius $r > 0$ and received at infinity is finite and diverges only as $r \to 0$.

To explore this further, we first plot the radial distance from the center as a function of the proper time of the infalling observer (Fig. 1a). The slope of this curve represents the velocity of the infalling object, and the free parameter $\beta$ is chosen within a specific range, which will be explained later. Notably, the observer reaches the center within a finite proper time and at a faster rate compared to the time required to reach a singularity in the Schwarzschild case.

In the Schwarzschild background, an infalling object appears to asymptotically approach the event horizon for a distant observer, taking an infinite amount of coordinate time to reach it, with the observed redshift diverging at the horizon. In contrast, for the BD solution, the redshift remains finite for all $r > 0$, and diverges only as the object approaches the naked singularity $r \to 0$ (Fig. 1b).

We can easily calculate the radius of the photon sphere by minimizing the effective potential energy of null geodesics, which results in the following expression,

$$r_\gamma = \sqrt{1-\beta^2}\,. \tag{8}$$

It is well known that objects without an event horizon can still possess a photon sphere, and hence a shadow feature [34,

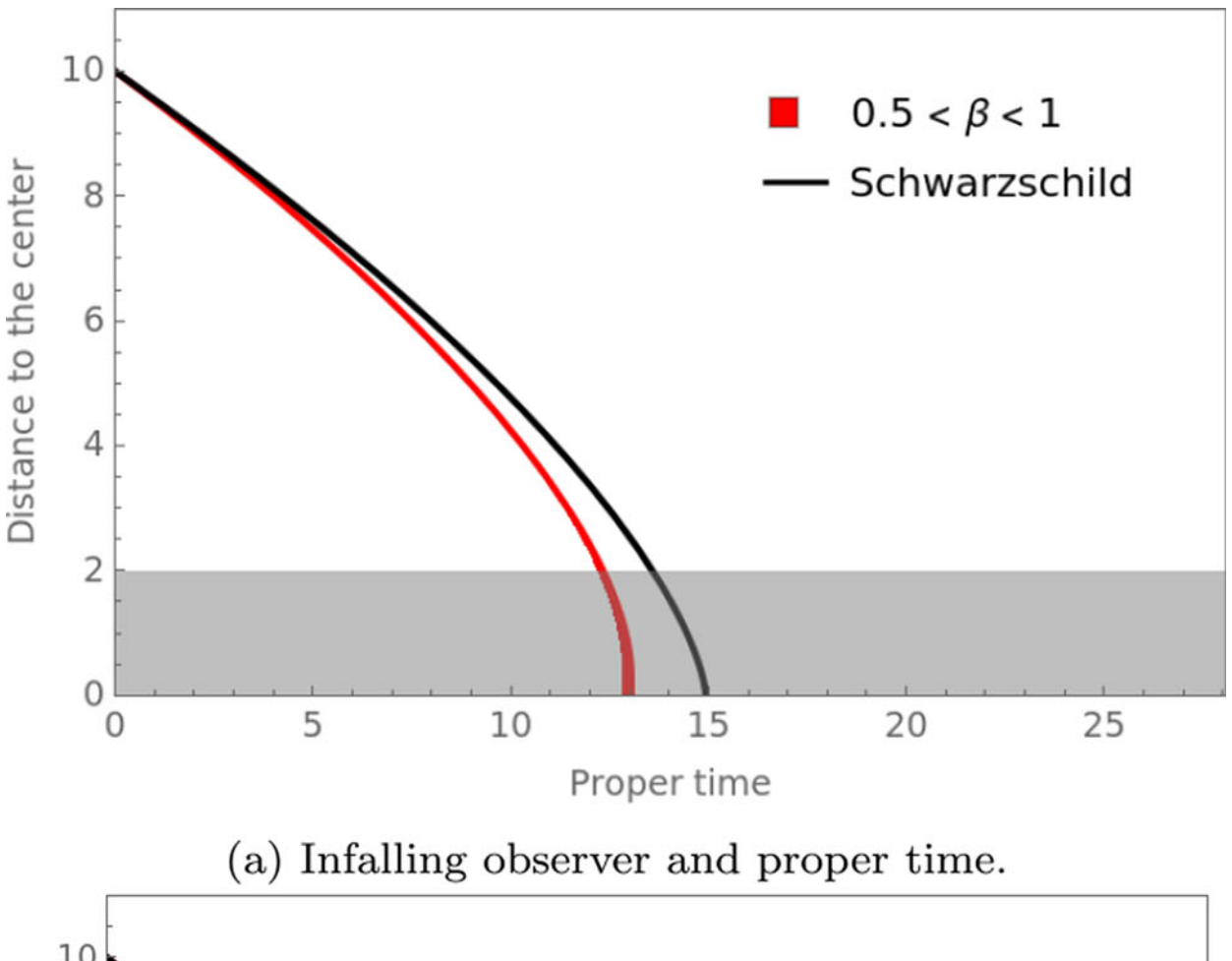

(a) Infalling observer and proper time.

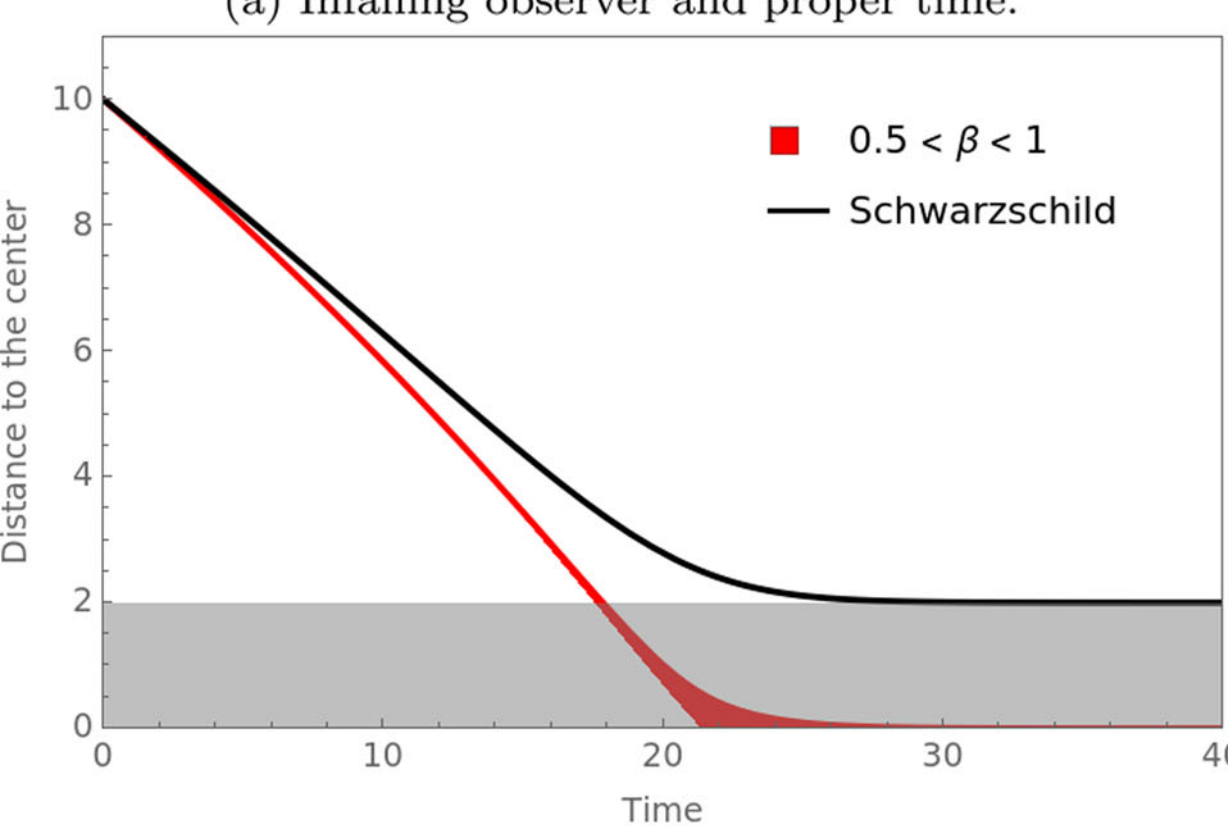

(b) Infalling matter as seen from a distant observer.

**Fig. 1** Radial motion with the initial conditions $r_{\text{in}} \equiv r(t=0) = 10$. We use $G = c = 1$ and set $M = 1$ for convenience here. The shaded region ($r < 2$) shows the $r$ values inside a Schwarzschild BH, shown here for comparison purposes. In plot **b**, the BD geodesics with the largest infalling time correspond to $\beta \to 0.5$

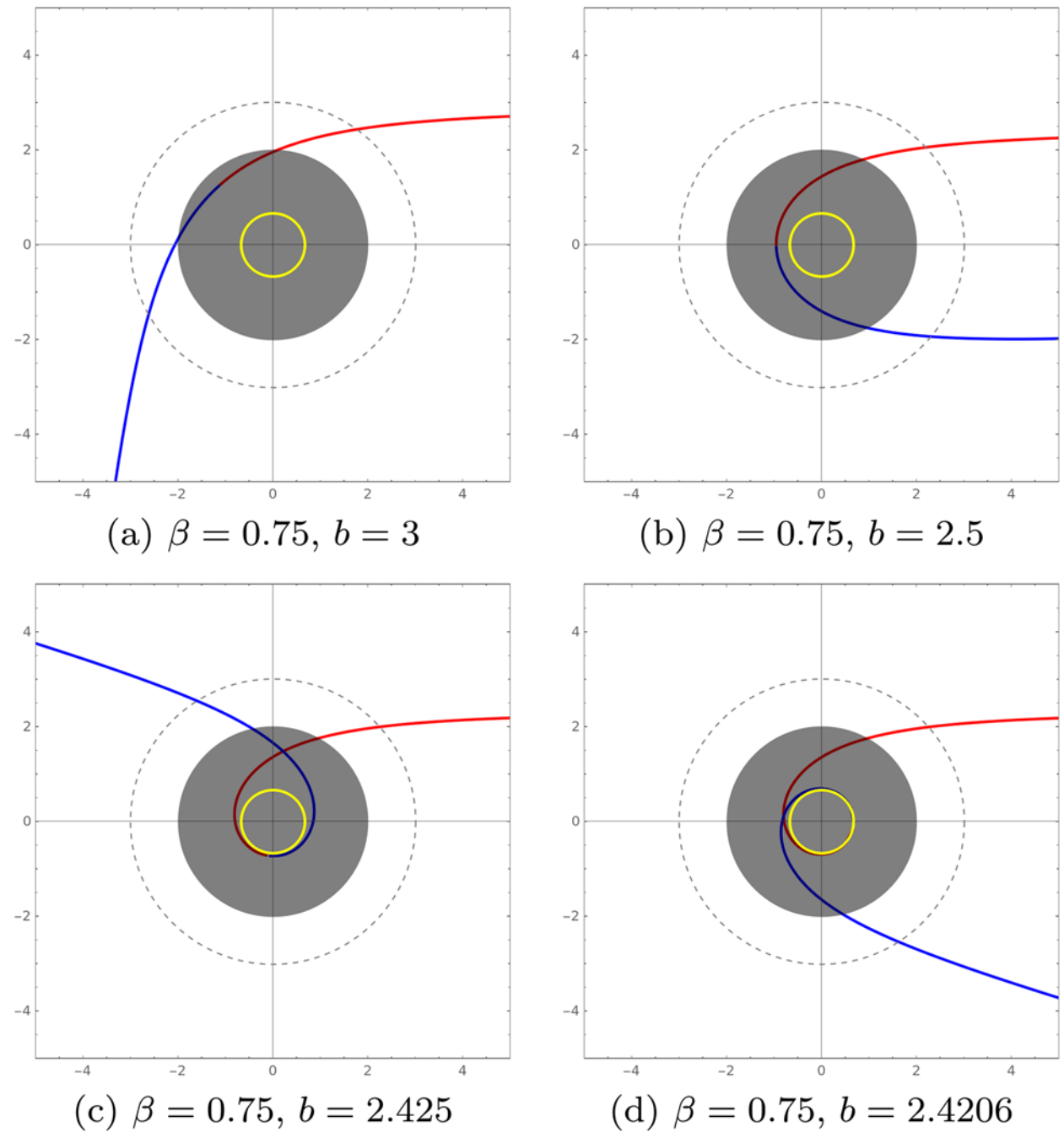

(a) $\beta = 0.75$, $b = 3$ (b) $\beta = 0.75$, $b = 2.5$ (c) $\beta = 0.75$, $b = 2.425$ (d) $\beta = 0.75$, $b = 2.4206$

**Fig. 2** Geodesics with 0, 1, 2, and 3 half-turns around the compact object. The red and blue curves represent the two solutions of the quadratic equation (6), which transform into each other at turning points. For all panels, we fix $\beta = 0.75$, with the Brans–Dicke photon ring ($r_\gamma = 0.661438$) shown in yellow. For comparison, the Schwarzschild horizon and its photon ring are displayed in gray. The impact parameter $b$ decreases from left to right and top to bottom, demonstrating the transition from unbound to increasingly wound trajectories

41–46]. In the case of the BD naked singularity, the photon sphere is described by the radius given in (8). This allows us to constrain the range of the model parameter $\beta$ for which a photon sphere exists.

1. *Existence of the photon sphere:* The condition that $r_\gamma$ must remain real imposes the constraint $\beta^2 < 1$ implying $\omega < 0$.
2. *Attractive gravity:* For gravity to remain attractive ($\omega > -2$), the parameter $\beta$ must be positive.
3. *Non-phantom fields:* To ensure the field $\psi$ is not a phantom or a ghost field ($\omega > -\frac{3}{2}$), we require $\beta > 0.5$.

Under realistic physical conditions, the BD parameter $\omega$ is constrained to the range $(-1.5, 0)$, which corresponds to $\beta \in (0.5, 1)$. Within this range, the photon ring lies entirely within half the Schwarzschild radius ($r_S = 2$) of a GR black hole with the same mass. As a result, the shadow of BD naked singularity is consistently smaller than that of a Schwarzschild black hole in GR.

To illustrate the photon trajectories we use equation (6). Since we have set the mass $M = 1$, the model has two free parameters – the model parameter $\beta$ and the initial impact parameter $b$. We set $\beta = 0.75$ and numerically compute the geodesics using (6) for various initial values $b_{\text{in}} \equiv b(r_{\text{in}})$ (we take $r_{\text{in}} = 1000$ – that makes the incoming trajectories parallel to the $x$ axis). Figure 2 shows the null trajectories with $m = 0, 1, 2, 3$ half turns respectively. For a particular choice of $\beta = 0.75$, we notice, as $m$ increases, $b$ goes to the critical impact parameter $b(r_\gamma(\beta)) = 2.4204$. This is the theoretical value of the radius of the shadow we expect from the simulated images or the observations.

For massive particles, the ISCO radius is determined using (5) and is given by $r_m = \sqrt{2}\sqrt{1-\beta^2 \pm \sqrt{1-\beta^2}}$. We choose $\beta = 0.75$. With the accretion disk peaking at the ISCO radius and illuminating the BD soliton, the resulting shadow size is $r_{\text{sh}} = 2.4204$. The axial image is shown in Fig. 3.

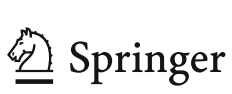

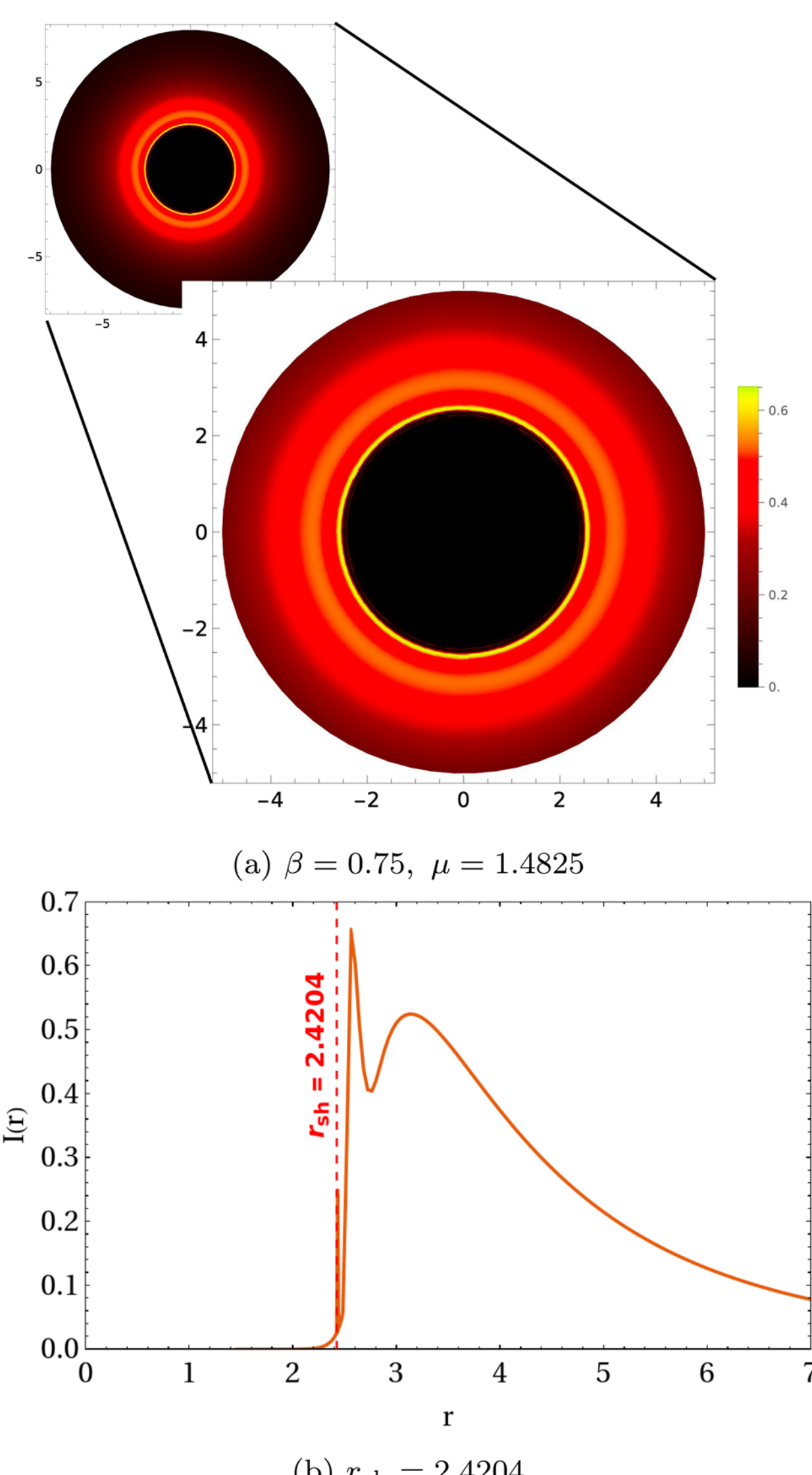

(a) $\beta = 0.75,\ \mu = 1.4825$

(b) $r_{\rm sh} = 2.4204$

**Fig. 3** The intensity profile parameters are set to $r_m = \mu = 1.4825$, and $\gamma = -2$ and $\sigma = 0.25$. The model parameter $\beta$ is fixed to 0.75. A very faint ring can be observed in the zoomed axial image of the BD naked singularity in **a** corresponding to the first intensity peak at the radius of the shadow $r_{\rm sh}$. This is a narrow peak formed by a few photons (of initially 2000 photons) making 2 half turns. The next peak is formed by the lensed photon trajectories which make one half turn. The last and the broader peak corresponds to those trajectories that are not sufficiently lensed to make half turns and correspond to the direct image of the ISCO

## 4 Discussion

Building on the works of [29,30], we investigate exact solutions within the BD theory. We note that the solution presented in [27] corresponds to a special case ($\omega = 0$) of the solution found in [29]. It describes naked singularity which has a photon sphere.

We restrict the BD parameter to a physically viable range, $-\frac{3}{2} < \omega < 0$, corresponding to $\beta^2 = \frac{\omega+2}{2} \in (0.25, 1)$. The photon sphere radius is then given by $r_\gamma = \sqrt{1-\beta^2}$. This range ensures attractive gravity and avoids ghost fields and faster-than-light propagation. For comparison, the value $\omega = -3/2$ is a special and largely studied value. Together with a potential, it corresponds to Palatini $f(R)$ gravity, which requires no screening mechanism to be compatible with Solar System data [47]. The case $\omega = -1$ corresponds to the dilaton case found from low-energy effective string theory which can be compatible with Solar System observations together with the Damour–Polyakov mechanism [48,49]. At last, it should be stressed that the limit of very large $\omega$ for BD is only necessary for the massless BD theory [50]. The results here discussed hold for any situation in which gravity can be locally approximated by a massless BD theory.

The Photon geodesics were computed numerically, and we derived the ISCO radius for massive particles. With the accretion disk peaked at the ISCO radius as an illumination source, we simulated the image of a BD naked singularity with $\beta = 0.75$, yielding a shadow size of $r_{\rm sh} = 2.4204$. Notably, BD naked singularity always has a smaller shadow than Schwarzschild black holes, given that both have the same mass.

The naked singularities with a photon sphere is the stately feature of this BD solution. Their compact size can exclude them from being the Galactic supermassive black hole. However, for M87*, the measurements from stellar dynamics ($6.6 \pm 0.4 \times 10^9 M_\odot$) [51] are almost twice the value inferred from a different technique using kinematics of ionized gas ($3.5^{+0.9}_{-0.7} \times 10^9 M_\odot$) (68%) [52] (see also [53]). This discrepancy means the BD naked singularity cannot be excluded for M87*, as the associated uncertainty in the mass translates into loose constraints on possible deviations of the shadow size from the standard Schwarzschild prediction.

The next-generation EHT (ngEHT) [54] will provide a modest improvement in angular resolution but may still be insufficient to resolve the $m = 1$ photon ring; however, incorporating parameterized modelling of the surrounding matter distribution could achieve effective "superresolution" [55]. Moreover, GRAVITY+ [56] will deliver better astrometric precision, tightening mass and distance constraints and enabling more decisive tests to distinguish BD naked singularities from standard black holes.

**Acknowledgements** PHP thanks João Luís Rosa and Gonzalo J. Olmo for their invaluable lectures on the ray-tracing code. This project greatly benefited from the iCOOP workshop, and the organizers are sincerely acknowledged. PHP also acknowledges *Fundação de Amparo à pesquisa e Inovação do Espírito Santo* (FAPES, Brazil) and *Coordenação de Aperfeiçoamento de Pessoal de Nível Superior* (CAPES, Brazil) for support. PHP and DCR thank the Institute of Theoretical Astrophysics (University of Oslo) for hospitality, where part of this work was developed. DCR also thanks *Centro Brasileiro de Pesquisas Físicas* (CBPF) and *Núcleo de Informação C&T e Biblioteca* (NIB/CBPF) for hospitality, where part of this work was done. He also acknowledges *Conselho Nacional de Desenvolvimento Científico e Tecnológico* (CNPq, Brazil), FAPES (Brazil) and *Fundação de Apoio ao Desenvolvimento da Computação Científica* (FACC, Brazil) for partial support. DFM thanks the Research Council of Norway for their support and the resources provided by UNINETT Sigma2 – the National Infrastructure for High-Performance Computing and Data Storage in Norway.

**Data Availability Statement** This manuscript has no associated data. [Authors' comment: Data sharing not applicable to this article as no datasets were generated or analysed during the current study.]

**Code Availability Statement** Code/software will be made available on reasonable request. [Authors' comment: The code generated during the current study is available from the corresponding author on reasonable request.]

Funded by SCOAP$^3$.

## Appendix A: Soliton solution

We can extend the calculations of [30] for general $\omega$ without restricting it to zero. We find the metric component $g_{rr} \equiv e^B$ takes the quadratic form,

$$e^B = 1 - 2(\omega + 2)Z^2 + \epsilon(Z - Z^2) . \tag{A.1}$$

Here, we recall that, $Z \equiv -\frac{\psi'}{2\psi}$ and the primes will indicate derivative *w.r.t* independent variable, $z \equiv \ln \frac{r}{r_*}$. There are two roots, $Z_1$, $Z_2$. For a particular choice of the constant $\epsilon = 2(\omega + 2)$, we find that, $Z_1 = \frac{1}{\sqrt{2(\omega+2)}} = -Z_2$.

The transition equation retains its form,

$$Z' = -Ze^B = -\epsilon Z(Z - Z_1)(Z + Z_1) . \tag{A.2}$$

However, now, the zero of the 'effective potential' on the *r.h.s*, $Z_1$ depends on the BD parameter $\omega$. The other zero is a trivial zero that can be identified with the Schwarzschild limit $Z \to 0$. The $tt$ component of the metric becomes,

$$e^A = \left( \frac{1 - \frac{Z}{Z_1}}{1 + \frac{Z}{Z_1}} \right)^{\frac{2}{\sqrt{\epsilon}}} . \tag{A.3}$$

We solve the transition equation to find,

$$\frac{Z}{Z_1} = \pm \frac{r}{\sqrt{1 - \left(\frac{r_*}{r}\right)^2}} .$$

For the choice, $\epsilon = 2(\omega + 2)$ we have the solution in (4). In [30], the solution for the particular choice of $\omega = 0$ was interpreted as a 'non-singular' soliton due to non-vanishing effective energy-momentum tensor density $\sqrt{-g}T^{\mu}_{\ \nu}$ of the BD scalar field. One has to note this argument cannot be relied upon as the tensor density is not a coordinate-independent quantity. Invariant curvature scalars like, the Kretschmann scalar can be shown to diverge for $r \to 0$ indicating a singularity (for $\omega > -2$). However interestingly these solution do have a soliton like behaviour. A more general solution and the physical aspects of such solutions will be discussed in future work.